\documentclass[12pt]{article}
\usepackage{epsf,amsmath,amssymb}

\newcommand\pubnumber{BNL-HET-01/2}
\newcommand\pubdate{February 2001}
\newcommand\hepnumber{hep-ph/0102266}

\def\csumb{HET, Physics Department\\
Brookhaven National Laboratory, Upton, NY 11973, U.S.A.\\
\tt\small rharlan@bnl.gov}
\def\support{\footnote{Supported by
{\it Deutsche Forschungsgemeinschaft}.}} 

\def\Title#1{\begin{center} {\Large\bf #1 } \end{center}}
\def\Author#1{\begin{center}{ \sc #1} \end{center}}
\def\Address#1{\begin{center}{ \it #1} \end{center}}

\newcommand\pubblock{\rightline{\begin{tabular}{l} \pubnumber\\
         \pubdate\\ \hepnumber \end{tabular}}}
\newenvironment{Abstract}{\begin{quotation}  }{\end{quotation}}
\newenvironment{Presented}{\begin{quotation} \begin{center} 
             Presented at the\end{center}
      \begin{center}\begin{large}}{\end{large}\end{center} \end{quotation}}

\makeatletter
\def\section{\@startsection{section}{0}{\z@}{5.5ex plus .5ex minus
 1.5ex}{2.3ex plus .2ex}{\large\bf}}
\def\subsection{\@startsection{subsection}{1}{\z@}{3.5ex plus .5ex minus
 1.5ex}{1.3ex plus .2ex}{\normalsize\bf}}
\def\subsubsection{\@startsection{subsubsection}{2}{\z@}{-3.5ex plus
-1ex minus  -.2ex}{2.3ex plus .2ex}{\normalsize\sl}}

\renewcommand{\@makecaption}[2]{%
   \vskip 10pt
   \setbox\@tempboxa\hbox{\small #1: #2}
   \ifdim \wd\@tempboxa >\hsize     % IF longer than one line:
       \small #1: #2\par          %   THEN set as ordinary paragraph.
     \else                        %   ELSE  center.
       \hbox to\hsize{\hfil\box\@tempboxa\hfil}
   \fi}

 \def\citenum#1{{\def\@cite##1##2{##1}\cite{#1}}}
\newcount\@tempcntc
\def\@citex[#1]#2{\if@filesw\immediate\write\@auxout{\string\citation{#2}}\fi
  \@tempcnta\z@\@tempcntb\m@ne\def\@citea{}\@cite{\@for\@citeb:=#2\do
    {\@ifundefined
       {b@\@citeb}{\@citeo\@tempcntb\m@ne\@citea\def\@citea{,}{\bf ?}\@warning
       {Citation `\@citeb' on page \thepage \space undefined}}%
    {\setbox\z@\hbox{\global\@tempcntc0\csname b@\@citeb\endcsname\relax}%
     \ifnum\@tempcntc=\z@ \@citeo\@tempcntb\m@ne
       \@citea\def\@citea{,}\hbox{\csname b@\@citeb\endcsname}%
     \else
      \advance\@tempcntb\@ne
      \ifnum\@tempcntb=\@tempcntc
      \else\advance\@tempcntb\m@ne\@citeo
      \@tempcnta\@tempcntc\@tempcntb\@tempcntc\fi\fi}}\@citeo}{#1}}
\def\@citeo{\ifnum\@tempcnta>\@tempcntb\else\@citea\def\@citea{,}%
  \ifnum\@tempcnta=\@tempcntb\the\@tempcnta\else
  {\advance\@tempcnta\@ne\ifnum\@tempcnta=\@tempcntb \else\def\@citea{--}\fi
    \advance\@tempcnta\m@ne\the\@tempcnta\@citea\the\@tempcntb}\fi\fi}
\makeatother

\newcommand{\api}{{\alpha_s\over \pi}}
\newcommand{\abbrev}{\small}
\newcommand{\msbar}{\overline{\mbox{\abbrev MS}}}
\newcommand{\code}{\tt}

\newcommand{\eqn}[1]{Eq.\,(\ref{#1})}
\newcommand{\fig}[1]{Fig.\,\ref{#1}}

\newcommand{\sct}[1]{Sect.\,\ref{#1}}
\newcommand{\dd}{{\rm d}}

\newcommand{\order}[1]{{\cal O}(#1)}
\newcommand{\bld}[1]{\boldmath{$#1$}}

\renewcommand{\Im}{{\rm Im}}
\begin{document}
\begin{titlepage}
\pubblock

\vfill
\def\thefootnote{\fnsymbol{footnote}}
\Title{Pad\'e approximation to fixed order QCD calculations}
\vfill
\Author{Robert V.~Harlander\support}
\Address{\csumb}
\vfill
\begin{Abstract}
  Pad\'e approximations appear to be a powerful tool to extend the
  validity range of expansions around certain kinematical limits and to
  combine expansions of different limits to a single interpolating
  function.  After a brief outline of the general method, we will review
  a number of recent applications and describe the modifications that
  have to be applied in each case. Among these applications are the
  $\msbar$/on-shell conversion factor for quark masses and the top decay
  rate at {\abbrev NNLO} in {\abbrev QCD}.
\end{Abstract}
\vfill
\begin{Presented}
5th International Symposium on Radiative Corrections \\ 
(RADCOR--2000) \\[4pt]
Carmel CA, USA, 11--15 September, 2000
\end{Presented}
\vfill
\end{titlepage}
\def\thefootnote{\arabic{footnote}}
\setcounter{footnote}{0}
%

%- }}}
%- {{{ intro:

\section{Introduction}
The field of radiative corrections in quantum field theory has always
been exciting and rapidly developing. At this conference various new
methods were presented that allow us to keep up with the ever increasing
complexity of the problems posed by modern particle physics.  Some of
these methods are concerned with the analytic evaluation of certain
classes of Feynman diagrams (see, e.g., \cite{gehrmann}).  Equally
important, however, is the development of systematic approximations for
complex problems. As demonstrated in various physical applications,
asymptotic expansions of Feynman diagrams prove to be a very efficient
tool for this purpose: they provide recipes to reduce the number of
dimensional scales (masses, momenta) that a diagram depends on (see,
e.g., \cite{smirnov}). The result is a series (possibly
asymptotic) in terms of ratios of these dimensional parameters.

However, the validity of an expansion is restricted to a certain -- in
general finite -- region of convergence. As we will see, Pad\'e
approximations have been used to enlarge the validity range of these
expansions. In some of the cases described below, expansions from
different limits could be combined to construct an interpolating
function which connects the individual, often non-overlapping regions of
convergence.\footnote{It might be appropriate to remark that Pad\'e
  approximations have also been used in the literature to estimate
  higher order corrections in perturbation theory (e.g.~\cite{aspade}).
  These considerations are of completely different nature than the
  fixed-order predictions which will be discussed in this talk.}

The outline of this review is as follows: we will begin by describing
the general method on the basis of the hadronic $R$ ratio.  This
quantity is extremely important in particle physics: not only is it
directly measurable at $e^+e^-$ colliders, but it also influences other
quantities, for example the running of the electro-magnetic coupling
constant $\alpha_{\rm QED}(s)$ or the anomalous magnetic moment of the
muon. The continuous efforts for an accurate evaluation of this quantity
have presently reached an accuracy of order $\alpha_s^3$, even though
only in the high energy limit.  At order $\alpha_s^2$, due to the
successful application of the Pad\'e procedure described below, the full
energy dependence is known.  At first, the method was applied to
non-singlet diagrams which clearly give the major contribution to $R$
(cf.\ \sct{pade}).  \sct{singlet} will describe the generalizations that
were necessary to evaluate the singlet contributions. Let us note that
recently also non-diagonal currents have been taken into account.  This
opens a new field of applications related to charged current reactions,
like single top production at hadron colliders.

The sections that follow are concerned with a different class of
applications of the Pad\'e procedure, namely the evaluation of on-shell
quantities.  First we describe a recent calculation of the conversion
factor from the quark mass in the $\msbar$ scheme to the on-shell scheme
at order $\alpha_s^3$. Due to the progress in the field of heavy quark
and top threshold physics, the evaluation of this factor was of utmost
importance.  The second application is the determination of {\abbrev
  NNLO} {\abbrev QCD} corrections to the top quark decay rate which is a
necessary input for a precise experimental determination of the top
quark properties at future colliders. Closely related is the evaluation
of the muon decay rate at second order in {\abbrev QED}, as well as
$\Gamma(b\to ue\bar\nu_e)$ to $\order{\alpha_s^2}$.  Each of the
on-shell quantities above has been calculated by two independent groups
with complementary methods. The agreement of the results once again
confirms the validity and accuracy of the Pad\'e method.

%- }}}
%- {{{ General procedure [pade]:

\section{General procedure}\label{pade}
We are not going to describe all the details of the procedure for
constructing Pad\'e approximants, because this has been done in the
literature to a sufficient extent (see in particular
\cite{BroFleTar,BaiBro,CheKueSteV,CheKueSteVASP,singlet,Kuehnradcor}).
Nevertheless, for the sake of a closed presentation, let us give the
main ideas by considering the by now ``classic'' example of the hadronic
$R$ ratio:
\begin{equation}
\begin{split}
R(s) = {\sigma(e^+e^-\to \mbox{hadrons})\over \sigma(e^+e^-\to
  \mu^+\mu^-)}\,.
\label{eq::rpi}
\end{split}
\end{equation}
The leading two orders in $\alpha_s$ for $R(s)$ are known in analytic
form. An analytic evaluation of the complete corrections at
$\order{\alpha_s^2}$ currently seems to be excluded. The exact answer is
known only for certain contributions, in particular the terms involving
a light fermion pair~\cite{HoaKueTeu}.

\begin{figure}
  \begin{center}
    \leavevmode
    \begin{tabular}{ccc}
      \epsfxsize=4.cm
      \epsffile[130 260 445 450]{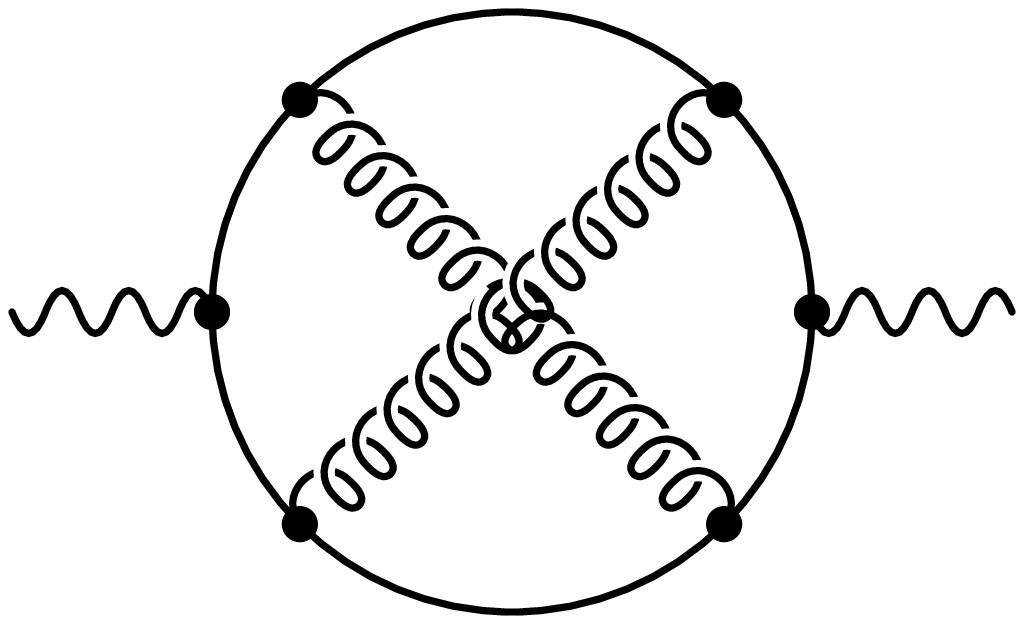} &
      \epsfxsize=4.cm
      \epsffile[130 260 445 450]{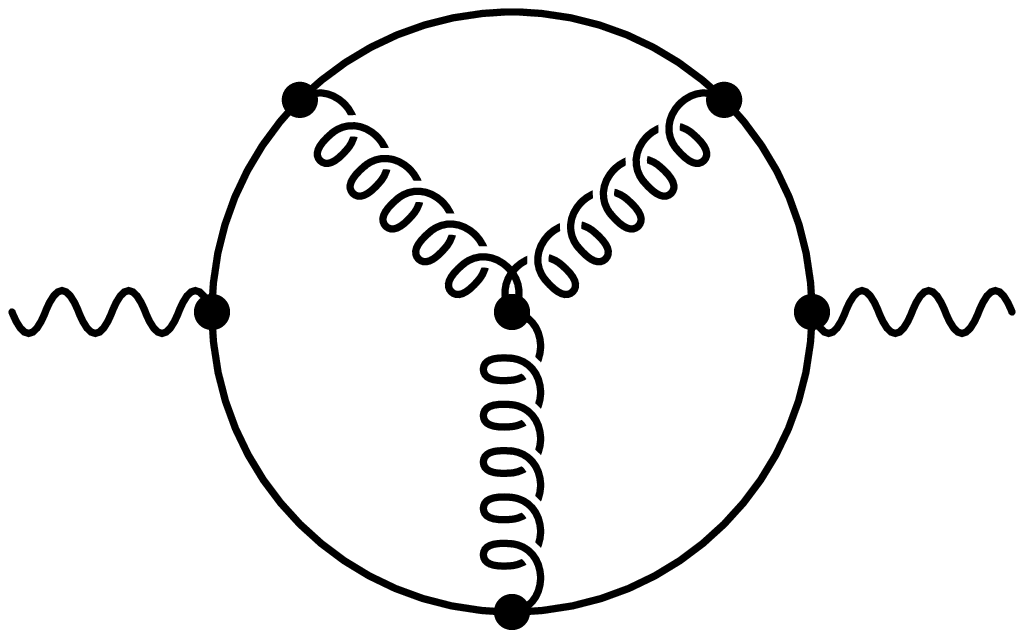} &
      \epsfxsize=4.8cm
      \epsffile[80 250 495 450]{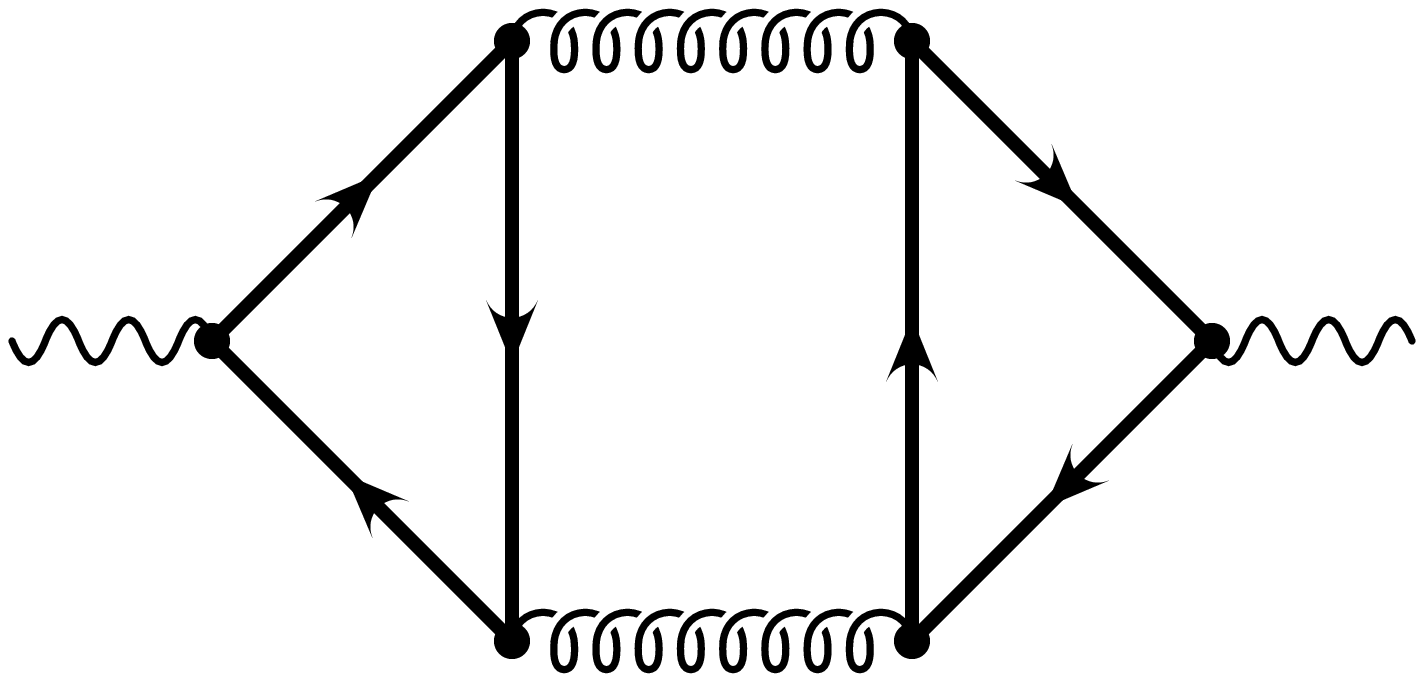} \\
      (a) & (b) & (c)
    \end{tabular}
    \parbox{14.cm}{
      \caption[]{\label{pidias}\sloppy
        Diagrams contributing to the polarization function $\Pi(z)$. The
        solid lines are quarks, the springy ones are gluons. The wavy
        lines represent the external currents.
        (a) and (b) are non-singlet, (c) is a singlet diagram.
        }}
  \end{center}
\end{figure}

For the other contributions one has to rely on approximations. There are
two obvious limiting cases for which this can be achieved: On the one
hand, if the center-of-mass energy is very large, one may set the quark
masses $m$ to zero. One may then use the equation
\begin{equation}
\begin{split}
R(s) = 12\pi\,\Im\Pi(s/(4m^2)+i0_+)
\end{split}
\end{equation}
which relates $R(s)$ to the imaginary part of the polarization function
$\Pi(z)$ along the upper branch of the cut $z\in[1,\infty]$ in the
complex plane. Sample diagrams for $\Pi(z)$ are shown in \fig{pidias}.
For the moment we will restrict the discussion to diagrams where the
external currents are connected by a single massive quark line
(\fig{pidias}\,(a) and (b)). Contributions where the external currents
are connected by a {\em massless} quark line and where the massive
quarks couple only to gluons (``gluon-splitting diagrams'') are
numerically unimportant and shall not be addressed here.  The
modifications for diagrams where each of the external currents is
connected to a separate Fermion line (``singlet diagrams'',
\fig{pidias}\,(c)) will be discussed in the next section.

Taking $m=0$ leads to massless propagator diagrams which can be
calculated using the integration-by-parts algorithm~\cite{ip} as
implemented in the {\code FORM} program~\cite{form} {\code
  MINCER}~\cite{mincer}.  One obtains
\begin{equation}
\begin{split}
  R(s) = 3\bigg\{ 1 + 
\api
  + \left(\api\right)^2\left[{365\over 24} - 11\zeta_3 +
    n_f\left(-{11\over 12} + {2\over 3}\zeta_3\right)\right]\bigg\}
  +\ldots\,,
\end{split}
\end{equation}
where the ellipse indicates higher order terms in $\alpha_s$ and in $m^2/s$.

On the other hand, the leading behavior in the opposite limit, where the
center-of-mass energy is close to threshold, i.e.\ $v\equiv
\sqrt{1-4m^2/s}\ll 1$, can be deduced from the Sommerfeld-Sakharov
formula and the two-loop result of the {\abbrev QCD} potential (for
details see \cite{CheKueSteV,CheKueSteVASP}). In general, these
considerations allow one to deduce the terms that are singular for $v\to
0$ (cf.\ Coulomb singularity) as well as the constant term.

\begin{figure}
  \begin{center}
    \leavevmode
    \begin{tabular}{cc}
      \epsfxsize=6.cm
      \epsffile[110 265 465 560]{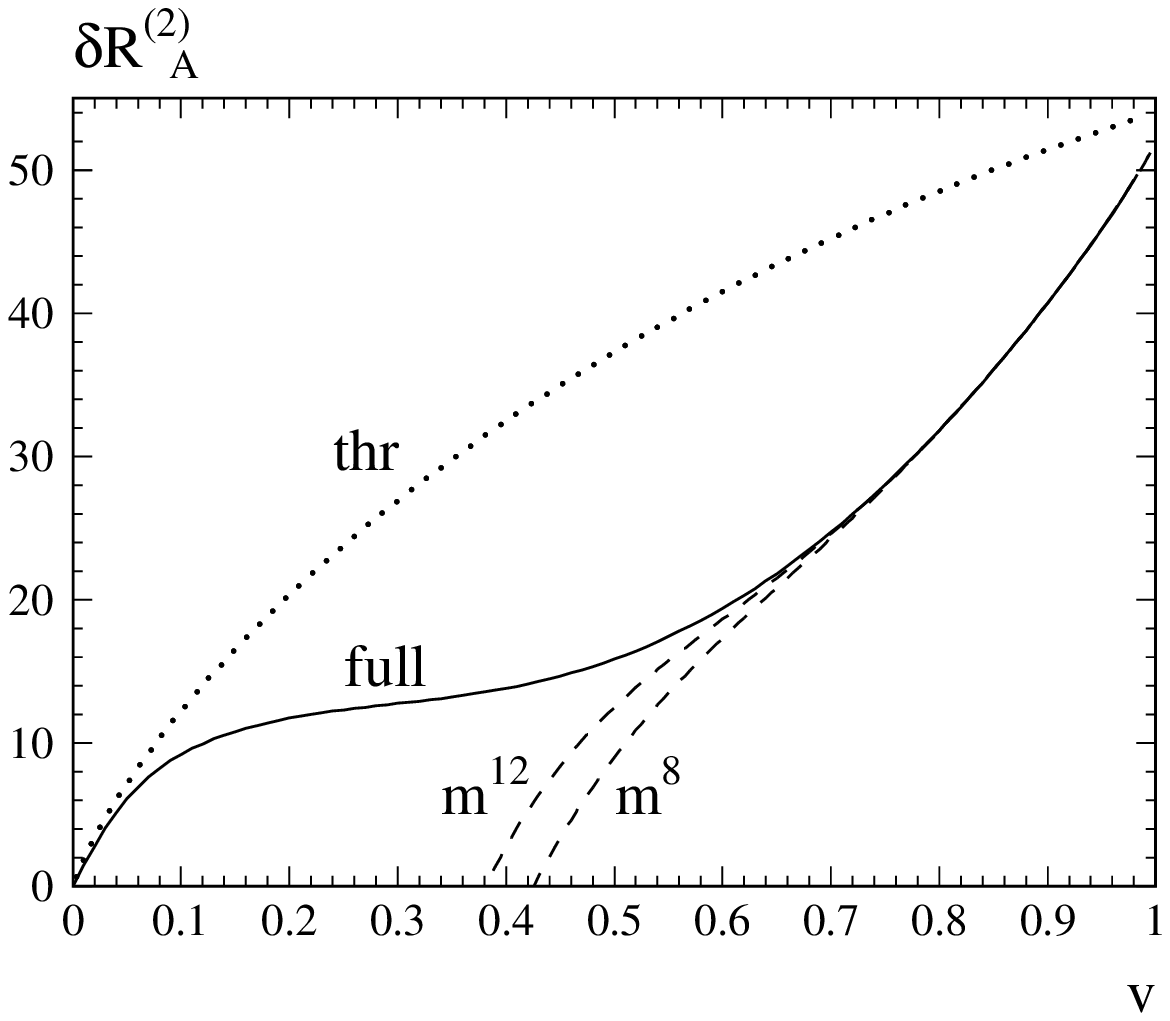} &
      \epsfxsize=6.cm
      \epsffile[110 265 465 560]{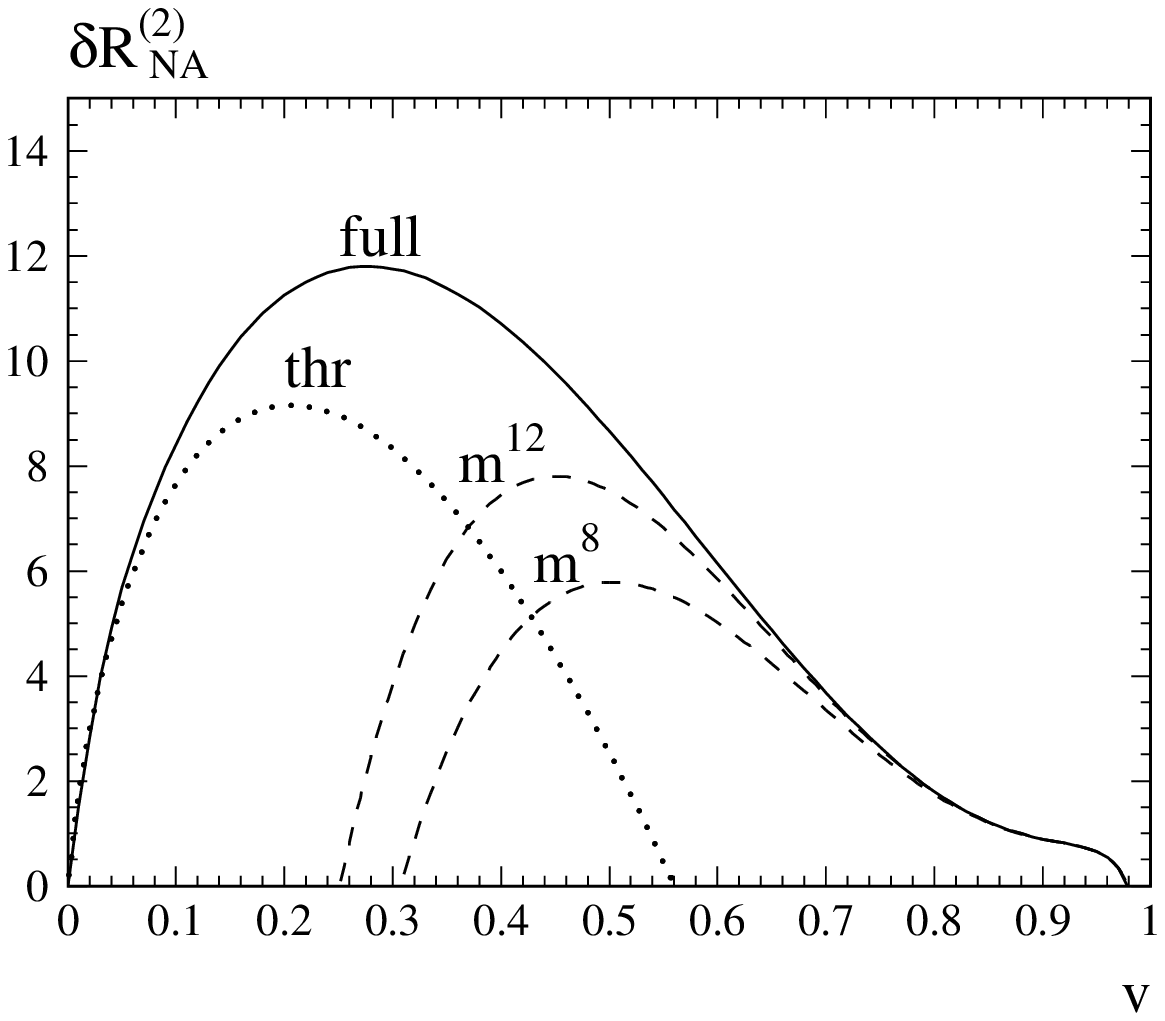}
    \end{tabular}
    \parbox{14.cm}{
      \caption[]{\label{fig::rvpadeexp}\sloppy
        Threshold expansion up to terms of order $v\ln v$~\cite{threxp}
        (dotted), high-energy expansion up to $m^{12}/s^6$~\cite{lmpapp}
        (dashed), and Pad\'e result~\cite{CheKueSteV} (solid). The
        singular and constant terms in $v$ have been subtracted and the
        renormalization scale is set to $\mu^2=m^2$.  Left: $C_{\rm
          F}^2$-term; Right: $C_{\rm F}C_{\rm A}$-term ($C_{\rm F}=4/3$
        and $C_{\rm A}=3$ are the Casimir operators of the fundamental
        and the adjoint representation of {\abbrev SU(3)}).}}
  \end{center}
\end{figure}

In both limits, however, one can do better. There are well-defined
methods to obtain {\it expansions} around the exact limits $m=0$ and
$s=4m^2$ (for reviews see \cite{smirnov,ustron}). These methods reduce
the original diagrams that depend on the two scales $m^2$ and $q^2$ to
single-scale integrals which can be solved analytically. In this way one
obtains approximations that are valid in certain ranges away from the
actual limits.  \fig{fig::rvpadeexp} shows the behavior of these
expansions as dashed and dotted lines. It immediately becomes clear that
their validity is restricted to a finite kinematical region.  It is the
purpose of this section to outline the procedure that leads to the {\em
  solid} lines in this figure, i.e.\ the construction of an
approximation which is valid over the full $v$ range.

The limits discussed above are the only two kinematically distinguished
points for $R(s)$ at $\order{\alpha_s^2}$ (we disregard the
four-particle threshold at $s=16m^2$ here).  Considering its connection
to the polarization function $\Pi(z)$, however, (cf.\ \eqn{eq::rpi}) there
clearly is another interesting point, namely $z=0$. The coefficients of
an expansion of $\Pi(z)$ around $z=0$ are called ``moments''. Because of
the cut at $z=1$, this expansion is expected to converge only for
$|z|<1$, and at first it seems to be impossible to extract any
information on $R(s)$ from it.  However, one has to recall that $\Pi(z)$
is analytic everywhere in the complex plane, except along the cut.  A
mapping of the form
\begin{equation}
\begin{split}
\omega = {1-\sqrt{1-z}\over 1+\sqrt{1-z}}
\label{eq::w2z}
\end{split}
\end{equation}
transforms the whole $z$-plane into the unit circle of the
$\omega$-plane, such that the upper (lower) branch of the cut gets mapped
to the upper (lower) semi-circle.  The points at $|z|=\infty$ go to
$\omega=-1$, while $z=0$ and $z=1$ correspond to $\omega=0$ and
$\omega=1$, respectively.

In order to arrive at a smooth function that is easy to approximate, one
subtracts the threshold singularities from $\Pi(z)$ (see above) as well
as the logarithms of the asymptotic high-energy expansion (this has to
be done in a suitable way in order not to generate logarithms for $z\to
0$).  After some more manipulations one arrives at a function
$P(\omega)$ whose value at $\omega=-1$ and whose first few Taylor
coefficients around $\omega=0$ are in one-to-one correspondence to the
expansion coefficients of $\Pi(z)$ around $z=0$ and $z=\infty$.

$P(\omega)$ is analytic within $|\omega|<1$ and thus its Taylor series
around $\omega=0$ converges inside this region.  However, we need the
values of $\Pi(z)$ {\em on} the cut, or equivalently, $P(|\omega|=1)$.
Thus one has to make sure that convergence of $P(\omega)$ is given not
only for $|\omega|<1$, but also for $|\omega|=1$.

This is the point where Pad\'e approximants come into play.
The definition of an $[n/m]$-Pad\'e approximant on a function $P(\omega)$ is
\begin{equation}
\begin{split}
P_{[n/m]}(\omega) = {a_0 + a_1 \omega + \cdots + a_n \omega^n\over 
  1 + b_1 \omega + \cdots + b_m \omega^m}\,.
\label{eq::padedef}
\end{split}
\end{equation}
It has been demonstrated in \cite{BroFleTar,BaiBro} that such an
approximant extends the convergence region of the Taylor expansion to
its border $|\omega|=1$.  Performing the mapping back to the $z$-plane,
and applying the inverse operations that led from $\Pi(z)$ to
$P(\omega)$, we thus have constructed a function that approximates
$\Pi(z)$ all over the complex plane, including the branches of the cut
from $z=1$ to $z=\infty$. The imaginary part of this function gives rise
to the solid lines in \fig{fig::rvpadeexp}.

A nice feature of Pad\'e approximation is that one has the freedom in
varying the parameters $n$ and $m$ in \eqn{eq::padedef}. If convergence
of the Pad\'e approximants was {\it not} given, this would manifest
itself in strong variations of the result for different values in $n$
and $m$.  For $\Pi(z)$, for example, this dependence is so weak that
different Pad\'e approximants produce curves that would be hardly
distinguishable from one another in \fig{fig::rvpadeexp} (see, e.g.,
\cite{Kuehnradcor}).

As it was mentioned above, this procedure was applied for the first time
in \cite{BaiBro} to the three-loop {\abbrev QED} vacuum polarization.
It was later on generalized to the {\abbrev QCD} case~\cite{CheKueSteV}
for various external currents~\cite{CheKueSteVASP} (vector,
axial-vector, scalar, pseudo-scalar).  In all of these papers, only the
leading two terms in the asymptotic expansion around $z\to \infty$ were
taken into account. After higher order terms in this limit became
available~\cite{lmpapp}, the method was extended to include
them~\cite{singlet} and a further stabilization of the Pad\'e
predictions was observed~\cite{rhdiss,Kuehnradcor}.
Recently~\cite{CheSteNonDiag} the method has been applied to
non-diagonal currents in order to derive the dominant contributions to
single top production at hadron colliders.

%- }}}
%- {{{ Singlet diagrams [singlet]:

\section{Pad\'e approximation for singlet diagrams}\label{singlet}
Above we described in some sense the ``optimal case'': information on
both the limits $z\to 0$ and $z\to \infty$ was available, and the
leading threshold behavior was known. Furthermore, the analytic
structure of $\Pi(z)$ was such that the expansion around $z=0$ had the
form of a plain Taylor expansion as opposed to an asymptotic series,
i.e., it did not contain any logarithms of $z$. This is due to the fact
that the discussion was restricted to the non-singlet contributions. The
corresponding diagrams (cf.~\fig{pidias}\,(a) and (b)) do not have
massless cuts: the external currents are connected by a single massive
quark line, and cutting the diagram in halves always involves a cut
through this line.

In the remaining part of this review we will be concerned with
exceptions to this ``optimal procedure.''  The first case we will
consider are the singlet contributions to $\Pi(z)$. They are
distinguished from the non-singlet diagrams in the sense that the
external currents are each connected to separate Fermion lines which in
turn are connected to each other by gluons (cf.\ \fig{pidias}\,(c)).  If
the external currents are vector-like, these diagrams vanish at
$\order{\alpha_s^2}$ due to Furry's theorem. If, however, one is
concerned with axial-vector, scalar, or pseudo-scalar currents, massless
cuts occur.\footnote{ In the axial-vector case, the purely gluonic cuts
  are zero according to the Landau-Yang theorem; but in order to avoid
  the axial anomaly, it is convenient to consider a full {\abbrev SU(2)}
  doublet like $(t,b)$. Taking $m_b=0$, the massless cuts arise from
  cuts involving $b$ quarks.}

These massless cuts spoil the analyticity of $P(\omega)$ (see
\sct{pade}) within $|\omega|<1$, and thus also the convergence of its
expansion around $\omega=0$.  Luckily, for the singlet diagrams the
analytic expressions of the massless cuts are known. Thus, denoting
these massless cuts as $R_{\rm ml}(s)$, we may employ the dispersion
relation to write
\begin{equation}
\begin{split}
  &\Pi_{\rm S}(z) = \Pi_{\rm ml}(z) + \hat
  \Pi(z)\,,\quad\mbox{with}\quad
  \hat \Pi(z) = C^{-1}\int_1^\infty\dd s{R_{\rm S}(s)\over s - 4m^2z}\\
  &\mbox{and}\quad \Pi_{\rm ml}(z) = C^{-1}\int_0^1\dd s{R_{\rm
      S}(s)\over s - 4m^2z} = C^{-1}\int_0^1\dd s{R_{\rm ml}(s)\over s
    - 4m^2 z}
\label{eq::piml}
\end{split}
\end{equation}
($C=12\pi$ for external vector and axial-vector currents, $C=8\pi$ for
scalar and pseudo-scalar currents).
Here, $\Pi_{\rm S}(z)$ is the singlet contribution to the polarization
function and 
\begin{equation}
\begin{split}
R_{\rm S}(s) = C\,\Im\Pi_{\rm S}(s/(4m^2) + i0_+)\,.
\end{split}
\end{equation}
$\hat \Pi(z)$ is analytic in the complex plane cut along
$z\in[1,\infty]$. It can be obtained by evaluating $\Pi_{\rm ml}(z)$
through \eqn{eq::piml} and subtracting it from $\Pi_{\rm S}(z)$, i.e.,
the result for the singlet diagrams (see \fig{pidias}~(c)). One can
then apply the Pad\'e procedure outlined in \sct{pade} to $\hat\Pi(z)$.

For details on the evaluation of the integral for $\Pi_{\rm ml}(z)$ in
\eqn{eq::piml} and the results we refer to~\cite{singlet}. At this point
it shall be sufficient to mention that the combination
of~\cite{CheKueSteVASP,HoaJezKueTeu} (non-singlet) and~\cite{singlet}
(singlet) provides the current knowledge of $R(s)$ at
$\order{\alpha_s^2}$. Some contributions are known analytically, and the
accuracy of the approximations in all the other cases is extremely good.
Let us also remark that at $\order{\alpha_s^3}$, $\Pi(z)$ is known in
its high-energy expansion up to the terms $\propto
m^4/s^2$~\cite{a3m0,a3m2,a3m4}. No moments for $z\to 0$ are available
yet, and therefore a Pad\'e approximation along the lines of the
previous section is still out of reach.

As another application of the methods described above let us note that
the results at $\order{\alpha_s^2}$ were combined with the one-loop
electro-weak corrections in order to predict the total cross section for
$e^+e^-\to t\bar t$ at a linear collider \cite{ewqcd}.

%- }}}
%- {{{ ms-bar/on-shell [msos]:

\section{Relation between \bld{\overline{\rm MS}} and on-shell quark
  mass}\label{msos}
The $\msbar$ scheme is a very convenient renormalization scheme, in
particular from the technical point of view. Renormalization constants
in the $\msbar$ scheme do not depend on any masses or momenta, which
means that their evaluation provides a certain freedom in choosing the
particular set of diagrams to calculate.  For example, the
four-loop quark anomalous dimension was computed using two completely
different approaches~\cite{gammam4}.

\begin{figure}
  \begin{center}
    \leavevmode
    \begin{tabular}{ccc}
      \epsfxsize=4.cm
      \epsffile[100 260 490 475]{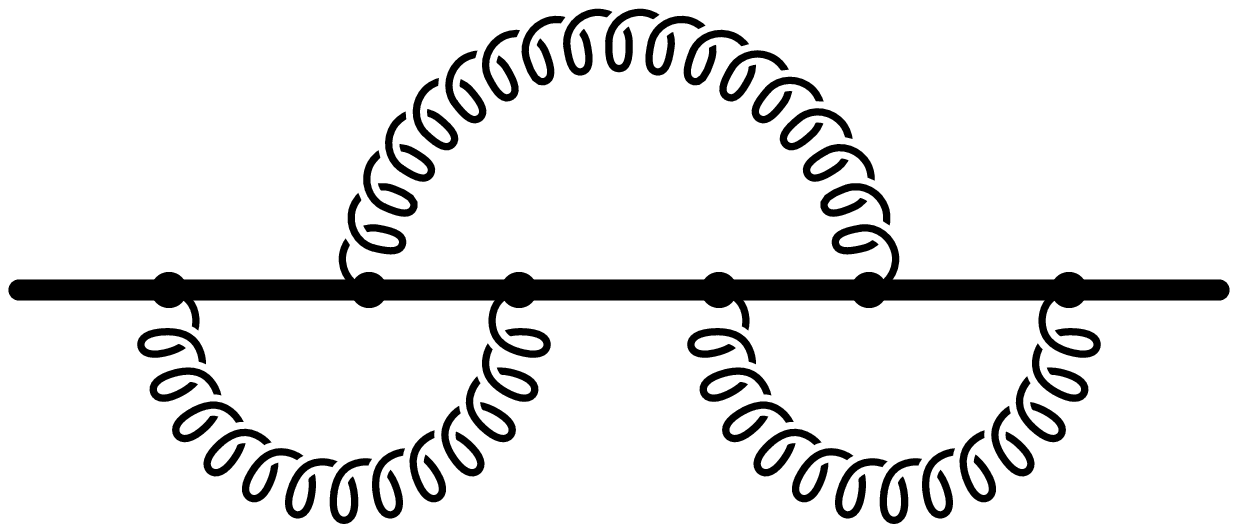} &
      \epsfxsize=4.cm
      \epsffile[100 260 490 475]{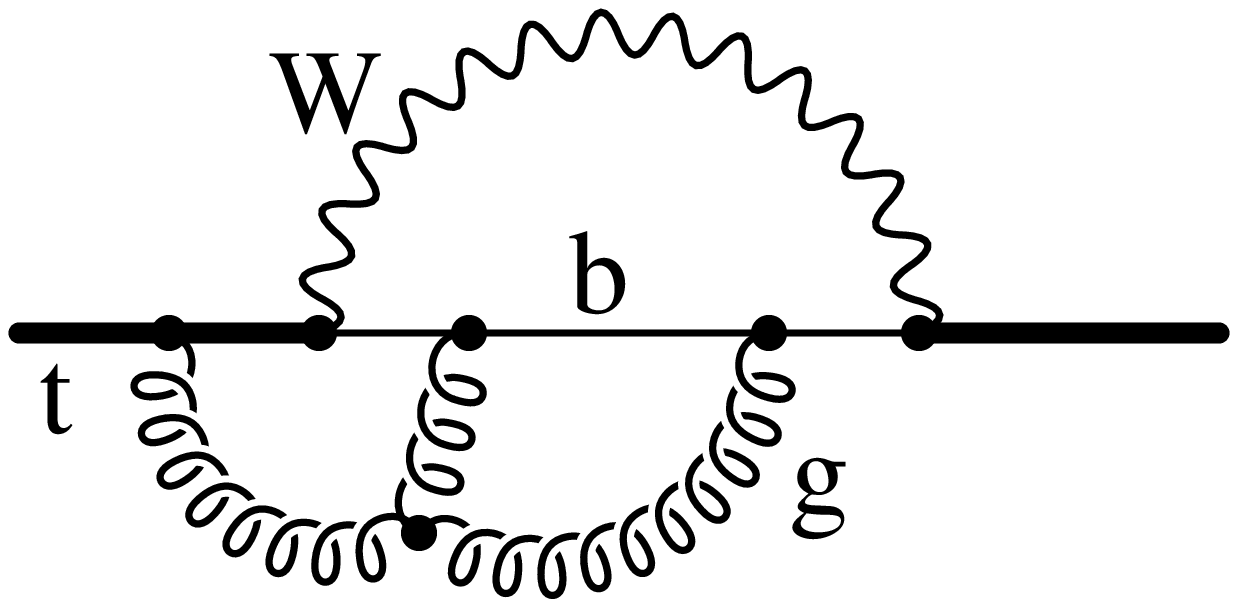} &
      \epsfxsize=4.cm
      \epsffile[145 476 485 650]{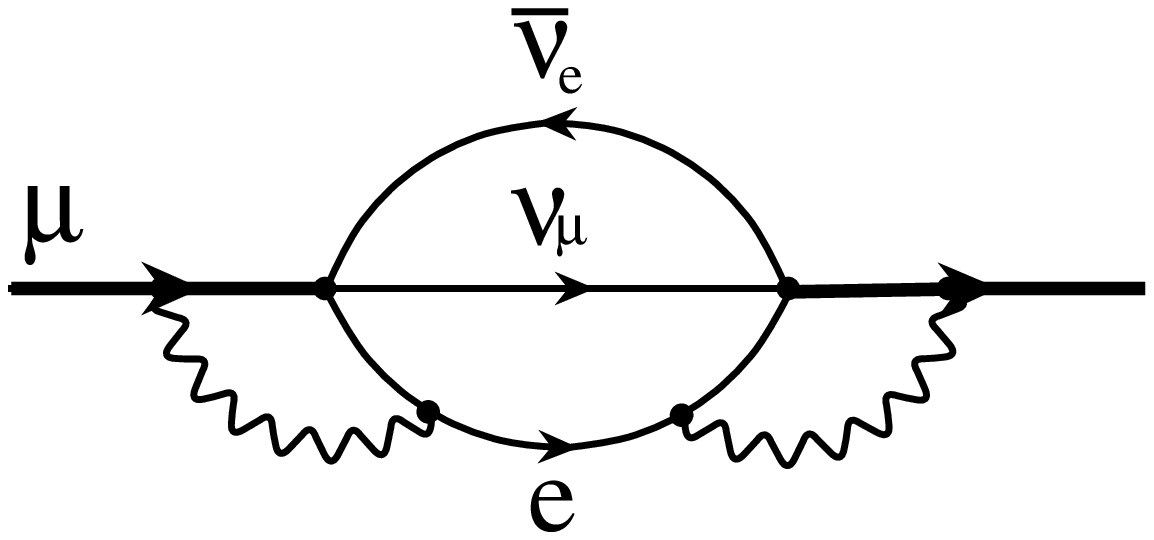}\\
      (a) & (b) & (c)
    \end{tabular}
    \parbox{14.cm}{
      \caption[]{\label{fprop}\sloppy
        Diagrams contributing to (a) $z_m$, (b) $\Gamma(t\to bW)$, and
        (c) $\Gamma(\mu\to e \bar\nu_e \nu_\mu)$.}}
  \end{center}
\end{figure}

However, comparison of the theoretical results to experimental data
often requires to express the involved quantities in terms of their
on-shell values.  This is why the conversion factor
\begin{equation}
\begin{split}
z_m = {M\over \bar m}
\end{split}
\end{equation}
that relates the on-shell to the $\msbar$ quark mass is of great
importance.  In order to obtain this quantity, one has to evaluate the
quark propagator
\begin{equation}
\begin{split}
\Sigma(q) = q\!\!\!/\,\Sigma_{\rm V}(q^2) + m\,\Sigma_{\rm S}(q^2)
\end{split}
\end{equation}
at the on-shell point $q^2=m^2$, where $q$ is the external momentum and
$m$ is the quark mass. A sample diagram that contributes to $\Sigma(q)$
at order $\alpha_s^3$ is shown in \fig{fprop}\,(a).

Technically, these diagrams carry only a single scale and should be
accessible through the integration-by-parts algorithm~\cite{ip}.
However, it turns out that the level of complexity for $n$-loop on-shell
diagrams is comparable to $(n+1)$-loop massive tadpole diagrams, for
example.  At the two-loop level, the problem of finding the recurrence
relations was solved in~\cite{msosa2}.

The three-loop case seemed to be out of reach for quite some time.  This
is why the first calculation of the conversion factor was performed
using a different, semi-analytic approach with the help of Pad\'e
approximants~\cite{msospade}.  Looking at the Fermion propagator as a
function of $z=q^2/m^2$ in the complex $z$ plane, we find a similar
structure as for the polarization function $\Pi(z)$ in \sct{pade}.
$\Sigma_{\rm S,V}(q^2)$ is analytic in the complex plane cut along $z\in
[1,\infty]$. Thus, in principle one can follow the same strategy as for
$\Pi(z)$: based on the expansions around $q^2=0$ and $q^2=\infty$ one
constructs an approximation for $\Sigma_{\rm S,V}(q^2)$ in the whole $q^2$
plane, including $q^2=m^2$ $(z=1)$, the point of interest.

\begin{figure}
  \begin{center}
    \leavevmode
    \begin{tabular}{c}
      \epsfxsize=14.cm
      \epsffile[100 570 510 725]{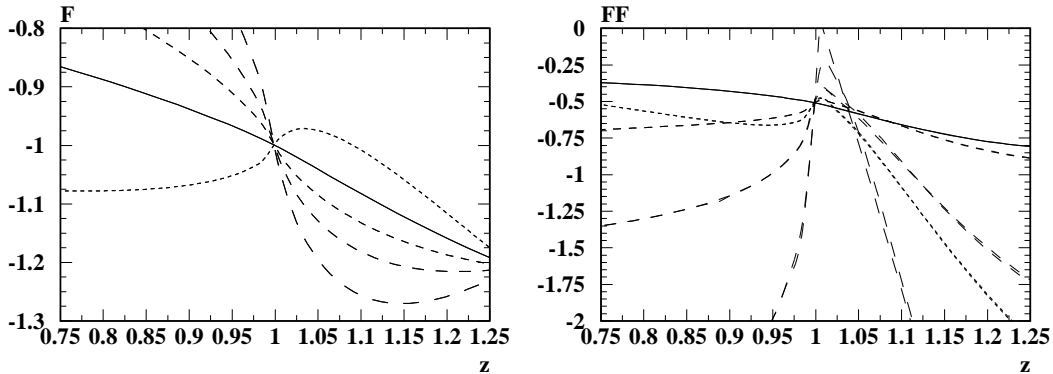}\\
    \end{tabular}
    \parbox{14.cm}{
      \caption[]{\label{xidep}\sloppy
        Gauge parameter dependence of $-g(z)$.  ($g(z)$ is constructed
        from $\Sigma_{\rm S,V}(q^2)$ in such a way that $g(1)=-z_m$ ---
        for details see \cite{msospade}). ``F'' denotes the
        $\order{\alpha_s}$ contribution, ``FF'' the $C_{\rm F}^2$ term
        at $\order{\alpha_s^2}$.  The solid line corresponds to
        $\xi=-2$, the others to $\xi=-5,0,+2,+5$ (from short to long
        dashes) [Figures taken from \cite{msospade}].  }}
  \end{center}
\end{figure}

A complication one has to face here is that the functions $\Sigma_{\rm
  S,V}(q^2)$ depend on the strong gauge parameter $\xi$ in
general.\footnote{We define the gluon propagator in $R_\xi$ gauge as
  $i(-g_{\mu\nu} + \xi q_\mu q_\nu/q^2)/(q^2+i0_+)$.}  Only at the
on-shell point $q^2=m^2$ are they gauge independent. Thus, the
expansions around $z=0$ and $z=\infty$ explicitly contain $\xi$. If they
could be re-summed exactly, $\xi$ would drop out in the limit $z\to 1$.
But here the exact re-summation will be replaced by a Pad\'e
approximation and thus the result depends on $\xi$ even for $z=1$.
However, the argumentation is that the dependence on $\xi$ is weak at
$z=1$, in the sense that any ``reasonable'' choice of $\xi$ leads to
valid predictions, with the spread among different choices being within
the error of the Pad\'e approximation. The question of what a
``reasonable'' choice is can be answered by making the natural
assumption that a smooth curve gets better reproduced by Pad\'e
approximants than a strongly varying one. Looking at \fig{xidep}, it is
not hard to decide that the favored values for $\xi$ are within a few
units around $\xi=0$.

Following the outlined procedure, the authors of \cite{msospade} were
able to deduce the value of $z_m$ with an uncertainty of around $\pm
3\%$.  Shortly after this result was presented, a second group published
the analytical result for the conversion factor $z_m$~\cite{MelRit}.
They managed to establish the recurrence relations for three-loop
on-shell diagrams derived by integration-by-parts, which provides an
extremely useful tool for various other applications
(see also~\cite{andrzej}).

%- }}}
%- {{{ top decay [topdecay]:

\section{Top decay to \bld{\alpha_s^2} and second order {\abbrev QED} 
  corrections to muon decay }\label{topdecay}
The decay rate of the top quark is expected to be measured at future
colliders with about 10\% accuracy. This is roughly the size of the
$\order{\alpha_s}$ corrections to this quantity, which is why the
evaluation of the $\alpha_s^2$ contribution was necessary.  One option
to evaluate this decay rate is to calculate the on-shell top quark
propagator to the appropriate order and to take the imaginary part. A
typical diagram whose imaginary part contributes to the rate
$\Gamma(t\to bW)$ at order $\alpha_s^2$ is shown in \fig{fprop}\,(b). In
a first approximation, one may take the bottom quark and the $W$ boson
to be massless. Then the lowest order diagram is a massless propagator.
Starting from $\order{\alpha_s}$, however, one is again faced with
proper on-shell diagrams. In contrast to the problem of the
$\msbar$/on-shell conversion factor (see \sct{msos}) the diagrams for
$t\to bW$ contain massless cuts due to the presence of the $b$ and the
$W$.  The solution of \cite{twb} (see also \cite{acmelproc}) to this
problem was to evaluate the expansion around $q^2/m_t^2=0$ and to take
the imaginary part {\em before} evaluating a Pad\'e approximation.  It
is therefore not possible to take information from the limit $q^2\to
\infty$ into account, because otherwise one receives contributions to
the imaginary part coming from the cut starting at $q^2=m_t^2$.

Nevertheless, the Pad\'e approximants constructed from the expansion
around $q^2=0$ alone give a fairly accurate result (judging from the
spread of the different approximants). In addition, it agrees well with
an earlier result~\cite{twbacmel} which relied on an expansion around
$1-m_b^2/m_t^2=0$. Considering the fact that these approaches are based
on expansions around two completely different limits -- which, in
addition, are both far from the physical point -- their agreement within
very small error bars is a clear demonstration of the power of the
applied methods.

Another example of this kind is the decay rate of the muon. Following a
strategy closely related to the one described above for top decay, it
was possible to evaluate the second order {\abbrev QED} corrections to
this quantity in a semi-analytical way~\cite{SeiSte}. Note, however,
that the corresponding diagrams contain four closed loops here (cf.\ 
\fig{fprop}\,(c)), even though one of them (made up by the neutrino
lines) is always a massless self-energy insertion.  Also in this case
the Pad\'e approximants agree nicely with the previously known
analytical result~\cite{RitStu} which provides an important check on the
latter.

Along the same lines one can also obtain the corrections of order
$\alpha_s^2$ to the decay rate $\Gamma(b\to ue\bar\nu_e)$, the only
technical difference being the presence of non-Abelian diagrams. Also
here the semi-analytical result~\cite{SeiSte} approximates the
analytical formula~\cite{timo} to high accuracy.

%- }}}
%- {{{ conclusions:

\section{Conclusions}
We reviewed the method and recent applications of Pad\'e approximation
to fixed order calculations in {\abbrev QCD}.  Originally developed for
the hadronic $e^+e^-$ cross section, the approach proved to be useful
also for a completely different class of problems related to on-shell
phenomena, for example in heavy quark physics.  Let us conclude by
pointing out that the continuously refining field of expansion
techniques for Feynman diagrams (see \cite{smirnov}) should pave the way
to numerous new applications for the Pad\'e method.

%- }}}
%- {{{ acknowledgments:

\paragraph*{Acknowledgments.}
I would like to thank the organizers of the {\abbrev RADCOR-2000}
symposium for the invitation, and all the participants for the pleasant
and fruitful atmosphere during the conference.  Travel support from the
High Energy Group at {\abbrev BNL}, {\abbrev DOE} contract number
{\abbrev DE-AC02-98CH10886}, is acknowledged.

%- }}}
%- {{{ bibliography:

\def\app#1#2#3{{\it Act.~Phys.~Pol.~}{\bf B #1} (#2) #3}
\def\apa#1#2#3{{\it Act.~Phys.~Austr.~}{\bf#1} (#2) #3}
\def\annphys#1#2#3{{\it Ann.~Phys.~}{\bf #1} (#2) #3}
\def\cmp#1#2#3{{\it Comm.~Math.~Phys.~}{\bf #1} (#2) #3}
\def\cpc#1#2#3{{\it Comp.~Phys.~Commun.~}{\bf #1} (#2) #3}
\def\epjc#1#2#3{{\it Eur.\ Phys.\ J.\ }{\bf C #1} (#2) #3}
\def\fortp#1#2#3{{\it Fortschr.~Phys.~}{\bf#1} (#2) #3}
\def\ijmpc#1#2#3{{\it Int.~J.~Mod.~Phys.~}{\bf C #1} (#2) #3}
\def\ijmpa#1#2#3{{\it Int.~J.~Mod.~Phys.~}{\bf A #1} (#2) #3}
\def\jcp#1#2#3{{\it J.~Comp.~Phys.~}{\bf #1} (#2) #3}
\def\jetp#1#2#3{{\it JETP~Lett.~}{\bf #1} (#2) #3}
\def\mpl#1#2#3{{\it Mod.~Phys.~Lett.~}{\bf A #1} (#2) #3}
\def\nima#1#2#3{{\it Nucl.~Inst.~Meth.~}{\bf A #1} (#2) #3}
\def\npb#1#2#3{{\it Nucl.~Phys.~}{\bf B #1} (#2) #3}
\def\nca#1#2#3{{\it Nuovo~Cim.~}{\bf #1A} (#2) #3}
\def\plb#1#2#3{{\it Phys.~Lett.~}{\bf B #1} (#2) #3}
\def\prc#1#2#3{{\it Phys.~Reports }{\bf #1} (#2) #3}
\def\prd#1#2#3{{\it Phys.~Rev.~}{\bf D #1} (#2) #3}
\def\pR#1#2#3{{\it Phys.~Rev.~}{\bf #1} (#2) #3}
\def\prl#1#2#3{{\it Phys.~Rev.~Lett.~}{\bf #1} (#2) #3}
\def\pr#1#2#3{{\it Phys.~Reports }{\bf #1} (#2) #3}
\def\ptp#1#2#3{{\it Prog.~Theor.~Phys.~}{\bf #1} (#2) #3}
\def\ppnp#1#2#3{{\it Prog.~Part.~Nucl.~Phys.~}{\bf #1} (#2) #3}
\def\sovnp#1#2#3{{\it Sov.~J.~Nucl.~Phys.~}{\bf #1} (#2) #3}
\def\sovus#1#2#3{{\it Sov.~Phys.~Usp.~}{\bf #1} (#2) #3}
\def\tmf#1#2#3{{\it Teor.~Mat.~Fiz.~}{\bf #1} (#2) #3}
\def\tmp#1#2#3{{\it Theor.~Math.~Phys.~}{\bf #1} (#2) #3}
\def\yadfiz#1#2#3{{\it Yad.~Fiz.~}{\bf #1} (#2) #3}
\def\zpc#1#2#3{{\it Z.~Phys.~}{\bf C #1} (#2) #3}
\def\ibid#1#2#3{{ibid.~}{\bf #1} (#2) #3}

\end{document}